\documentstyle[preprint,aps,epsf,floats]{revtex} 

\def\GeV{\ \rm{ GeV} }

\begin{document}

\preprint{\tighten \vbox{\hbox{} }}

\title{The Resummed Rate for $B\rightarrow X_s\gamma$}

\author{Adam K.\ Leibovich, and I.\ Z.\ Rothstein}

\address{
Department of Physics,
Carnegie Mellon University,
Pittsburgh, PA 15213}

\maketitle

{\tighten
\begin{abstract}

In this paper we investigate the effect of the resummation of
threshold logs on the rate for $B\rightarrow X_s\gamma$. We calculate
the differential rate $d\Gamma/dE_\gamma$ including the infinite set
of terms of the form $\alpha_s^n \log^{n+1}(1-x)$ and $\alpha_s^n
\log^n(1-x)$ in the Sudakov exponent. The resummation is potentially
important since these logs turn into $\log(2E_{cut}/m_b)$, when the
rate is integrated from the lower cut $x=2E_{cut}/m_b$ to 1.  The
resummed rate is then convolved with models for the structure function
to study whether or not the logs will be enhanced due to the fermi
motion of the heavy quark.  A detailed discussion of the accuracy of
the calculation with and without the inclusion of the non-perturbative
effects dictated by the $B$ meson structure function is given.  We
also investigate the first moment with respect to $(1-x)$, which can
be used to measure $\bar\Lambda$ and $\lambda_1$.  It is shown that
there are some two loop corrections which are just as large as the
$\alpha_s^2 \beta_0$ term, which are usually expected to dominate.  We
conclude that, for the present energy cut, the threshold logs do not
form a dominant sub-series and therefore their resummation is
unnecessary. Thus, the prospects for predicting the rate for
$B\rightarrow X_s\gamma$ accurately, given the present energy cut, are
promising.

\end{abstract}
}


\newpage 

\section{Introduction}
The process $B\to X_s\gamma$ is considered fertile ground for
discovering physics beyond the standard model due to the fact that its
leading order contribution is a one loop effect.  Of course, any hope
of finding new physics is predicated on our ability to control the
theoretical prediction within the standard model.  This decay rate can
be calculated in a systematic expansion in $\alpha_s$, $1/(M_w,m_t)$
and $1/m_b$. Tremendous effort has gone into calculating the decay
rate at next to leading order in the strong coupling.  The calculation
is broken into three stages. First the full standard model is matched
onto the four fermion theory.  It is then run down to the scale $m_b$,
after which the inclusive decay rate may be calculated in heavy quark
effective field theory \cite{MW}.

It is in this last step that certain effects arise which jeopardize
the accuracy of the calculation. In particular, due to the fact
that there is an imposed energy cut on the photon, a new small parameter
enters the calculation, namely, the distance to threshold
$1-2E_\gamma/m_b$. As $E_\gamma$ approaches $m_b/2$, we begin to probe
larger and larger distances, until finally, uncontrollable non-perturbative
effects dominate.

The subject of this paper will center on the threshold logs,
$\log(1-2E_{cut}/m_b)$, which become parametrically large as $E_{cut}$
approaches the end point. As is well known, these logs appear because
the limited phase space obstructs the KLN cancellation of the IR
sensitivity. Thus for large enough values of $E_{cut}$ the
perturbative expansion breaks down, and must be resummed.  The leading
double logs resum into the well known Sudakov exponent. However, the
resummation beyond leading log accuracy is much more complicated.  The
next to leading logs were resummed in Ref.~\cite{AR} in a systematic
fashion\footnote{We take the opportunity to correct several typos in
\cite{AR}. After Eq.~(40) the definition of $B$ should include a
factor of $C_F$. In Eq.~(57) both $\gamma$ and $B$ should have plus
signs in front of them and in Eq.~(74) $g^{(1)}$ should be replaced by
$\gamma^{(1)}$.  In Eq.~(75) the argument of the exponential should be
$\log N g_1+g_2$.} utilizing the moment space factorization
\cite{SK}. However, since the result was given in moment space, an
analysis of the result in terms of measurable quantities is
difficult. Indeed, in Ref.~\cite{AR} it was found that in moment space
the sub-leading logs in the exponent actually dominate the leading
logs. Thus, it is important that we make sure that the physical
prediction not suffer the same consequences.  Previously in the
literature the subleading logs in $x$ space ($x$ being the variable
$2E_\gamma/m_b$) were naively estimated by exponentiating the one loop
subleading log \cite{AG,KN}.  In Ref.~\cite{DSU} an attempt was made
at including the subleading logs in a non-trivial fashion, however,  
it was not shown that all the logs of a given order were summed.

Here, we will build upon the results in Ref.~\cite{AR} and present the
resummed differential rate at next to leading log accuracy in $x$
space.  We then integrate the rate as a function of the energy cut to
get the rate to be compared with experiment.  Then the effects of
the fermi motion,  which in principle could
enhance the threshold logs, are included in this resummation. 
Finally, the net theoretical errors
incurred due the the unknown shape of the structure function
responsible for the fermi motion are estimated.

\section{The one loop result}

By combining the operator product expansion (OPE) and the heavy quark
effective theory (HQET), it is possible to calculate the decay
spectrum for inclusive heavy meson decays in a systematic expansion in
$\alpha_s$ and $\Lambda_{\rm QCD}/m_b$ \cite{CGG}.  In the endpoint
region, $E_\gamma \to m_b/2$, both the perturbative and $\Lambda_{\rm
QCD}/m_b$ expansions break down.  The non-perturbative corrections can
be resummed into a structure function \cite{N,BSUV}, which will be
discussed in Section \ref{CWTSF}.

The calculation of inclusive $B$ decay rates begins with the
low-energy effective Hamiltonian
\begin{equation}
H_{eff} = -\frac{4G_F}{\sqrt{2}}V_{ts}^*V_{tb}
   \sum_{i=1}^8C_i(\mu)O_i(\mu),
\end{equation}
where $G_F$ is the Fermi constant, $V_{ij}$ are elements of the
Cabibbo-Kobayashi-Maskawa matrix, $C_i(\mu)$ are Wilson coefficients
evaluated at a subtraction point $\mu$, and $O_i(\mu)$ are dimension
six operators.  For $B\to X_s\gamma$, the only operators that give a
relevant contribution are
\begin{eqnarray}
O_2 &=& (\bar{c}_{L\alpha}\gamma^\mu b_{L\alpha})
          (\bar{s}_{L\beta}\gamma_\mu c_{L\beta}), \nonumber\\
O_7 &=& \frac{e}{16\pi^2}m_b
      \bar{s}_{L\alpha}\sigma_{\mu\nu}F^{\mu\nu}b_{R\alpha}, \\
O_8 &=& \frac{g}{16\pi^2}m_b \bar{s}_{L\alpha}
   \sigma_{\mu\nu}G_a^{\mu\nu}T^a_{\alpha\beta}b_{R\beta}. \nonumber
\end{eqnarray}
Here $e$ is the electromagnetic coupling, $g$ is the strong coupling,
$m_b$ is the $b$ quark mass, $F^{\mu\nu}$ is the electromagnetic field
strength tensor, $G_a^{\mu\nu}$ is the strong interaction field
strength tensor, and $T^a$ is a color $SU(3)$ generator.  Near the
endpoint, the decay rate is dominated by the $O_7$ operator.  In
particular, the decay rate due to the $O_7$ operator is
\begin{equation}
\label{onelooprate}
\left. \frac1{\Gamma_0}\frac{d\Gamma}{dx}\right|_{x<1} = 
   \frac{(2x^2-3x-6)x + 2(x^2-3)\log(1-x)}{3(1-x)},
\end{equation}
where $x=2E_\gamma/m_b$ and 
\begin{equation}
\Gamma_0 = \frac{G_F^2|V_{ts}^*V_{tb}|^2\alpha\,C_7^2\,m_b^5}{32\pi^4}.
\end{equation}
As $x\to1$ this contribution diverges as $\log(1-x)/(1-x)$.  The
integrated rate is finite due to virtual corrections at $x=1$.  Only
decays with large photon energies can be detected, due to large
background cuts with the current experimental cut on the photon energy given by
$E_\gamma>2.1\GeV$ \cite{CLEO}.

\section{The Systematics of the Expansion}
The perturbative series near the endpoint schematically takes the form
\begin{eqnarray}
\frac{d\Gamma}{dx}\, =&& C_{11}\alpha_s\frac{\log(1-x)}{1-x} 
  + C_{12}\alpha_s\frac1{1-x} + C_{13}\alpha_s \nonumber\\
&+& C_{21}\alpha_s^2\frac{\log^3(1-x)}{1-x}
  + C_{22}\alpha_s^2\frac{\log^2(1-x)}{1-x}
  + C_{23}\alpha_s^2\frac{\log(1-x)}{1-x} + \cdots \nonumber\\
&+& C_{31}\alpha_s^3\frac{\log^5(1-x)}{1-x} 
  + C_{32}\alpha_s^3\frac{\log^4(1-x)}{1-x}
  + C_{33}\alpha_s^3\frac{\log^3(1-x)}{1-x} + \cdots \nonumber\\
&+& \cdots.
\end{eqnarray}
All non-integrable functions here are tacitly assumed to be ``plus''
distributions.  When we integrate form $\delta$ to 1, then each power
of $\log^{n-1}(1-x)/(1-x) $ turns into $\log^n(1-\delta)$.  Given this
series we may ask, how close to the endpoint can we study the spectrum
and still expect to get the right answer?  Clearly, if $\alpha_s
\log^2(1-\delta) \ll 1$ we may simply truncate the series at the first
term. To go closer to the endpoint, {\it i.e.}~larger values of $\delta$,
it would seem that we must sum the complete lower triangle along a
diagonal, a daunting task.  For instance, if we just sum the first
column, then we can not let $\alpha_s \log^2(1-\delta) \simeq 1$ since
at some point the terms in the next column may grow.  However, this
naive criteria is incorrect. The reason for this is that the series is
known to exponentiate into a particular form. In fact, the series
resums into a function of the form
\begin{equation}
\label{sudseries}
\log[\Gamma(\delta)] = \log(1-\delta) g_1[\alpha_s \log(1-\delta)]
+g_2[\alpha_s \log(1-\delta)] + \alpha_s g_3[\alpha_s \log(1-\delta)]+\cdots.
\end{equation}
This result implies that the aforementioned triangle assumes a
definite structure. The resummation does {\it not} sum the entire
triangle.  There will always be cross terms between higher order and
lower order terms in the exponent, which arise in the expansion of the
exponential, that have not been kept. What the resummed form does
tell you though, is that these cross terms can be neglected.  The important
point to notice about this structure is that if we truncate the series
in Eq.~(\ref{sudseries}), and assume that the last term we kept is
$O(1)$, then there is no ``growth'' in higher orders in $\alpha_s$ as
there is in the case of organizing the calculation in terms of the
columns.  

If we keep the first term in the expansion of $g_1$ we get the usual
Sudakov double logarithm. In that case $\delta$ must satisfy the
condition $\alpha_s^2 \log^3(1-\delta) < 1$. Note, that this
does not allow $\alpha_s \log^2(1-\delta)$ to become arbitrarily large
(practically this will not be an issue).  In general, then if we
expand $g_1$ up to $O(\alpha_s^{n-1})$, the requirement becomes
$[\alpha_s \log(1-\delta)]^n \log(1-\delta)< 1$. Thus, once
$\alpha_s \log(1-\delta)$ approaches one, we must for all
intensive purposes include the entire $g_1$, as well as $g_2$.

On a practical note it is clear that we can not let the logs become
arbitrarily large.  As $\delta$ approaches one we begin to probe
momenta of order $\Lambda_{\rm QCD}$, where the perturbative approach
stops making sense.  Equivalently, we are asking questions about
details of the hadronization process once we reach the resonance
regime. Presently, the experimental limit on $\delta$ is $2
(2.1)/m_b\simeq 0.88$.\footnote{Here $x$ is defined in terms of the
quark mass, this will be modified when we include the effects of the
structure function.}  Thus $\alpha_s \log^2(1-\delta) \simeq .95$, if
we do not include the factor of $1/\pi$ which accompanies each power
of $\alpha_s$.  If we include the factor of $\pi$, then it may well be
that we have not resummed the {\it dominant} piece of the series, and
resumming non-logarithmic terms could be just as important.  The
question of the inclusion of a factor of $\pi$ is a dicey numerical
issue given the size of coefficients in the expansion.  
Thus it is necessary to perform the resummation to determine whether
or not the logs form a dominant sub-series.

\section{The Resummation}

The resummation is performed in moment space, where the rate
factorizes into a short distance hard part $(H_N)$, a soft part
$(S_N)$ and a jet function $(J_N)$.  Using renormalization group
techniques it is possible to show that \cite{Sterman}
\begin{equation}
\label{resummed}
\sigma_NJ_N = \exp\left[\log N g_1(\chi) + g_2(\chi)\right],
\end{equation}
where $\chi = \alpha_s(m_b^2)\beta_0\log N$, and
\begin{eqnarray}
g_1 &=& -\frac{C_F}{2\pi\beta_0\chi}
      [(1-2\chi)\log(1-2\chi) - 2(1-\chi)\log(1-\chi)], \\
g_2 &=& -\frac{C_F k}{4\pi^2\beta_0^2}
            \left[2\log(1-\chi)-\log(1-2\chi)\right]
   -\frac{C_F\beta_1}{2\pi\beta_0^3}\left[\log(1-2\chi)-2\log(1-\chi) 
\frac{}{}\right.\nonumber\\
&& + \left.\,\frac12\log^2(1-2\chi)-\log^2(1-\chi)\right] 
   -\frac{3C_F}{4\pi\beta_0}\log(1-\chi) 
   -\frac{C_F}{2\pi\beta_0}\log(1-2\chi) \nonumber\\
&& + \, \frac{C_F\gamma_E}{\pi\beta_0}
    \left[\log(1-2\chi) - \log(1-\chi)\right].
\end{eqnarray}
In the above, $\beta_0 = (11\,C_A - 2\,N_f)/(12\pi)$, $\beta_1 = (17\,C_A^2
-5\,C_A\,N_f-3\,C_F\,N_f)/(24\pi^2)$, and $k = C_A\,(67/18-\pi^2/6) -
10\,T_R\,N_f/9$.

To go back to $x$-space, we must take the inverse-Mellin transform of
Eq.~(\ref{resummed}) which is given by
\begin{equation}
\label{xresum}
\frac{d\Gamma}{dx} = \frac1{2\pi i}\int_{C-i\infty}^{C+i\infty}M_Nx^{-N}dN
\end{equation}
Then we use the identity \cite{CMNT}
\begin{equation}
\label{invmell}
\frac{1}{2\pi i} \int_{C-i\infty}^{C+i\infty}x^{-N} 
e^{\log(N)\,F[\alpha_s \log(N)]}dN = -x\frac{d}{dx} \left\{
\Theta (1-x) e^{l\,F(\alpha_s\,l)} \times\left[1+E(\alpha_s,l)\right]\right\}
\end{equation}
where
\begin{equation}E(\alpha_s,l)=\sum_{k=1}^{\infty}\alpha_s^k \sum_{j=0}^k
e_{kj}l_j
\end{equation} 
represents subleading log contributions and 
\begin{equation}
l=-\log\left(-\log(x)\right).
\end{equation}
It is important at this point to note a crucial difference between
this calculation and resummations carried out in threshold production.
In threshold production (Drell-Yan for instance) the overall sign of
$g_1$ is flipped (in standard schemes). This arises as a consequence
of the fact that the Sudakov suppression in the parton distribution
function overwhelms the analogous suppression in the hard scattering
amplitude.  Thus, after subtraction, the overall sign of the exponent
changes. This difference in sign changes the nature of the inverse
Mellin transform. In particular, note that in the case of a negative
exponent (our case) the inverse Mellin transform is integrable even if
we ignore the derivative acting on the step function, which is
multiplying a function that is manifestly zero.  Whereas in the case
of the positive sign, the $\Theta$-function is crucial to define the
inverse Mellin transform in a distribution sense. The expansion of the
series for the Drell-Yan case has a non-integrable pole on the
positive axis in the Borel plane.  Which is to say that using the
above approximation has introduce subleading terms which nonetheless
become numerically important due to spurious factorial growth.  Thus,
as pointed out by Catani {\it et.~al.}~to avoid any large
corrections/ambiguities one should not use Eq.~(\ref{invmell}) but
instead one must perform the inverse Mellin transform exactly
\cite{CMNT}.  In our case, the expansion of the analytic result also
leads to a factorially divergent series, but it is sign alternating
and therefore Borel summable.  More simply put, the $x$ space form is
integrable.

To calculate the inverse Mellin transform with next-to-leading log
accuracy ({\it i.e.}~sum all logs of the form
$\alpha_s^n\log^n(1-\delta)$ in the exponent) we will perform the
integral numerically. In so doing we we choose the integration contour
such that the constant $C$ above is chosen to lie to the left of the
Landau pole singularity and to the right of all other potential poles.

\section {Effects on the Total Rate and First Moment}

Let us now consider the effects of the resummation on the partonic
rate. We will start by expanding the function $g_1$ in $\alpha_s$ and
systematically improving the approximation by including more
terms. The expansion of $g_1$ leads to a convergent series with a
radius of expansion given by $x_{max}=1-\exp[-1/(2\alpha_s\beta_0)]$.
To integrate the rate from $\delta$, we use the fact that the first
moment is one, and thus instead integrate from 0 to $\delta$, thus
avoiding the region where our approximation of the moments breaks
down.  If we keep up to $O(\alpha_s^n$) in this expansion, then the
terms we drop are $O(\alpha_s^{n+1} \log^{n+2}(1-\delta))$. Thus
keeping more terms in the expansion allows us to take $\alpha_s
\log(1-\delta)$ closer to one.  Figure 1 shows the percentage
contribution to the total rate stemming from the resummation, as a
function of the photonic energy cut. We see that as we include more
terms in the expansion of $g_1$ in the exponent, the expansion
converges to a fixed rate. Also, in Fig.~1, we show the rate which
includes all the subleading logs of the form
$[\alpha\log(1-\delta)]^n$ in the exponent, which is calculated by
keeping the full form of $g_1$ and $g_2$ and performing the inverse
Mellin transform numerically.  We see that overall the effect of
resummation is small. Moreover, the subleading corrections are of the
same size as the leading corrections. This is {\it NOT} because the
expansion is ill behaved, but because the subleading terms are not
truly suppressed compare to the terms of order $\alpha_s^n
\log^{n+1}(1-\delta)$, simply because the logs is not that large. We
do not have a dominant sub-series to sum.  The inclusion of $g_2$
gives a larger result because the $O(\alpha_s)$ piece in $g_2$ has a
larger coefficient than the $O(\alpha_s)$ piece in $g_1$.
\begin{figure}[t]
\centerline{\epsfysize=11truecm  \epsfbox{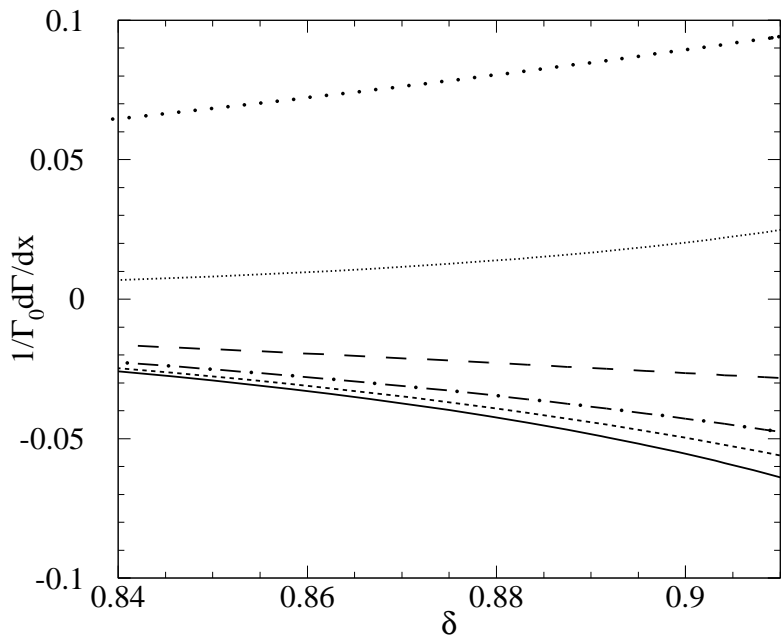} }
\tighten{
\caption[]{\it The effects of resummation on the rate, integrated from
$\delta$ to $1$.  The solid curve is the leading log result.  The
sparse dotted curve is the numerically inverted next to leading log
result.  The tight dotted curve is the rate expanding $g_1$ to
$O(\alpha_s)$.  The long dashed curve is expanding $g_1$ to
$O(\alpha_s^2)$, the dot-dashed curve is expanding $g_1$ to
$O(\alpha_s^3)$, and the short dashed curve is expanding $g_1$ to
$O(\alpha_s^4)$.}}
\end{figure}

Thus, the resummation is effectively just including some piece of the
two loop result. This can be clearly seen in Fig.~2 where
we show the resummed double log rate with the $O(\alpha_s)$ piece
subtracted out, and the $O(\alpha_s^2)$ piece from the expansion of the
double log resummation. We see that the results are nearly identical.
Also in Fig.~2 we show the $O(\alpha_s^2)$ derived from expanding
the full result, including $g_1$ and $g_2$, performing the inverse
Mellin transform exactly using
\begin{equation}
\frac{1}{2\pi i} \int_{C-i\infty}^{C+i\infty}x^{-N} 
\log^n(N)dN =  -x \frac{d}{dx} 
\left[\left(-\frac{d}{dk}\right)^n
\left\{
\Theta(1-x) e^{(k-1)\log[\log(1/x)] -  \log[\Gamma(k)]}
\right\}\right]_{k\to1}.\nonumber
\end{equation}
We see that the $O(\alpha_s^2)$ piece coming from formally subleading
log terms is actually larger than those coming from the leading logs
terms. This was hinted at in the results of the moment space
calculation in Ref.~\cite{AR}. However, this is not because the expansion
for the rate is converging poorly, as the effect is still small compared
to the lower order contribution, but because the logs are just not
that large.
\begin{figure}[t]
\centerline{\epsfysize=11truecm  \epsfbox{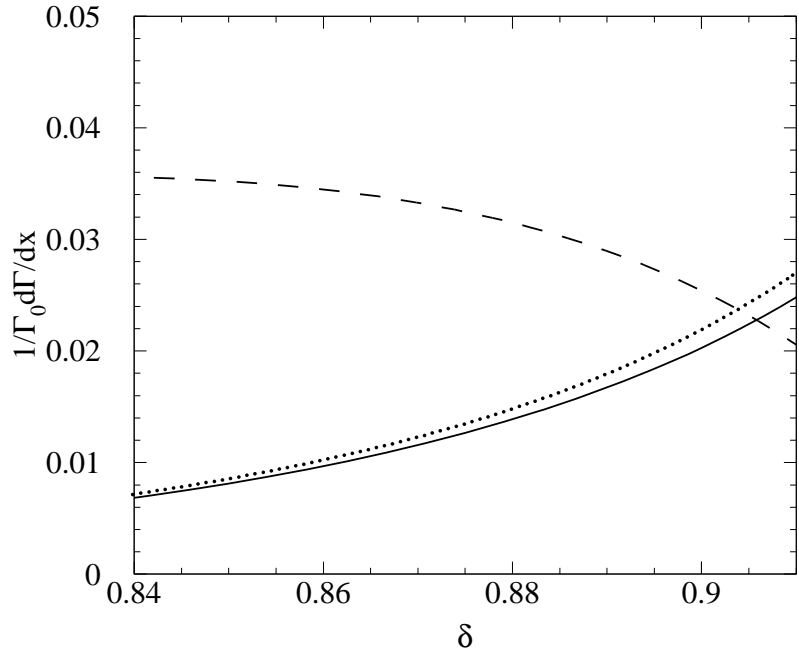} }
\tighten{
\caption[]{\it  The solid curve is the resummed double log rate, with
the $O(\alpha_s)$ piece subtracted out.  The dotted curve is the
$O(\alpha_s^2)$ piece of the resummed double log rate.  The dashed
curve is the $O(\alpha_s^2)$ piece of the full resummed rate,
including both $g_1$ and $g_2$.}}
\end{figure}

We may use also our results to pick out certain terms in the two loop
calculation. In particular, we may expand the exponent and determine
the coefficients of the terms of order $\alpha_s^2 \log^3(1-x)/(1-x)$
and $\alpha_s^2 \log^2(1-x)/(1-x)$.  Lower order logs will not be
correctly reproduced given the fact that we have dropped terms of
order $\alpha_s$ in the exponent. Expanding Eq.~(\ref{resummed}) to
$O(\alpha_s^2)$ we find the terms
\begin{equation}
\label{blm}
\left.\frac{d\Gamma}{dx}\right|_{O(\alpha_s^2)} = 
  \frac{8\alpha_s^2}{9\pi^2} \frac{\log^2(1-x)}{1-x}+
  \frac{\alpha_s^2}{\pi^2} \left(2\beta_0 \pi + \frac{14}{3}\right)
  \frac{\log(1-x)}{1-x}.
\end{equation}
Recently, the BLM correction to the first moment of $d\Gamma/dx$ has
been calculated.  This correction, proportional to
$\alpha_s^2\beta_0$, typically is a good approximation to the full two
loop result.  The interest in calculating the corrections to the first
moment is to get a better handle on the extraction of the HQET
parameters $\bar{\Lambda}$ and $\lambda_1$. If we compare the piece
proportional to $\beta_0$ in Eq.~(\ref{blm}), we find agreement with
the corresponding piece found in Ref.~\cite{LLMW}\footnote{Our
definition differs for $\beta_0$ differs by a factor of $4\pi$ from
\cite{LLMW}.}.  However, we also note that the contribution of the
second term in Eq.~(\ref{blm}), not proportional to $\beta_0$, is just
as big as the BLM corrections calculated in Ref.~\cite{LLMW}. This
does not rule out the possibility of a cancellation with other terms
such that the BLM result in the two loop result still
dominates. However, it does put into question the issue of BLM
dominance, and calls out for the complete two loop calculation of the
first moment.

\section{Convolution with the Structure Function}
\label{CWTSF}
As is well known, as one probes the spectrum closer the end point, the
OPE breaks down, and the leading twist non-perturbative corrections
must be resummed into the $B$ meson structure function
\cite{N,BSUV}. Formally we may write the light cone distribution
function for the heavy quark inside the meson as
\begin{equation}
f(k_+)=\langle B(v)\mid \bar{h}_v \delta(k_+-iD_+) h_v \mid B(v)\rangle.
\end{equation}
While the shape of this function is unknown, 
the first few moments of $f(k_+)$,
\begin{eqnarray}
A_n &=& \int dk_+ k_+^n f(k_+) \nonumber\\
&=& \langle B(v)\mid \bar{h}_v \delta(iD_+)^n h_v \mid B(v)\rangle,
\end{eqnarray}
are known; $A_0=1$, $A_1=0$, and $A_2=-\lambda_1/3$.
$f(k_+)$ has support over the range $-\infty<k_+<\bar{\Lambda}$.
It has been asserted that the support below $-\bar{\Lambda}$ dies off
exponentially \cite{MN}, but no formal proof has been given.

The effects of the fermi motion of the heavy quark can be included by
convoluting the above structure function with the differential rate,
\begin{equation}
\frac{d\Gamma}{dE_\gamma} = \int_{2E_\gamma-m_b}^{\bar\Lambda}
dk_+ f(k_+) \frac{d\Gamma_p}{dE_\gamma}(m_b^*),
\end{equation}
where $d\Gamma_p/dE_\gamma$ is the rate Eq.~(\ref{onelooprate}), or
the resummed version Eq.~(\ref{xresum}), written as a function of the
``effective mass'' $m_b^* = m_b + k_+$, {\it i.e.}, $x =
2E_\gamma/m_b^* = M_Bx_B/m_b^*$.  The new differential rate is now a
function of $x_B=2E_\gamma/M_B$.  The addition of the structure
function resums the leading-twist corrections and moves the endpoint
of the spectrum from $E_\gamma=m_b/2$ to the physical endpoint
$E_\gamma=M_B/2$.\footnote{The true end point of course takes
into account the final state masses.}
The cut rate may then be written as 
\begin{equation}
\label{cutratestruc}
\Gamma_H\left[\frac{2 E_{cut}}{M_B}\right] =
  \int_{2 E_{cut}-m_b}^{\bar{\Lambda}}dk_+\,\Gamma_p
  \left[\frac{2 E_{cut}}{m_b+k_+}\right]
\end{equation}
where $\Gamma_p [2 E_{cut}/(m_b+k_+)]$ is the partonic rate with a cut
at $x_p=2 E_{cut}/(m_b+k_+)$. From this result we can see that at
smaller values of $k_+$ the infra-red logs will be enhanced because
the phase space available to gluon emission is curtailed.  On the
other hand, the structure function will be suppressed at values of
$k_+<-\bar{\Lambda}$.  Thus, we expect any enhancement to be vitiated
by the effects of the structure function.

We will use following two different ans\"atze for the shape of the
structure function \cite{KN}:
\begin{eqnarray}
F_1(k_+) &=& N_1(1-k_+/\bar\Lambda)^a e^{(1+a)k_+/\bar\Lambda},\label{SF1}\\
F_2(k_+) &=& N_2(1-k_+/\bar\Lambda)^b e^{-c(1-k_+/\bar\Lambda)^2}\label{SF2}.
\end{eqnarray}
The parameters $N_i,a,b$, and $c$ can be determined from the known
moments $A_{0,1,2}$. Note that both of these forms have exponential
suppression for  $k_+<-\bar{\Lambda}$, which need not be true.
The values of $\bar{\Lambda}$ and $\lambda_1$
have been extracted from the data and are highly correlated.
We will use the central values $(\bar{\Lambda}=0.39,\lambda_1=-0.19)$ 
and the one sigma values  $(\bar{\Lambda}=0.28,\lambda_1=-0.09)$ 
and  $(\bar{\Lambda}=0.50,\lambda_1=-0.09)$ determined in 
\cite{GKLW}.

\begin{figure}[t]
\centerline{\epsfysize=11truecm  \epsfbox{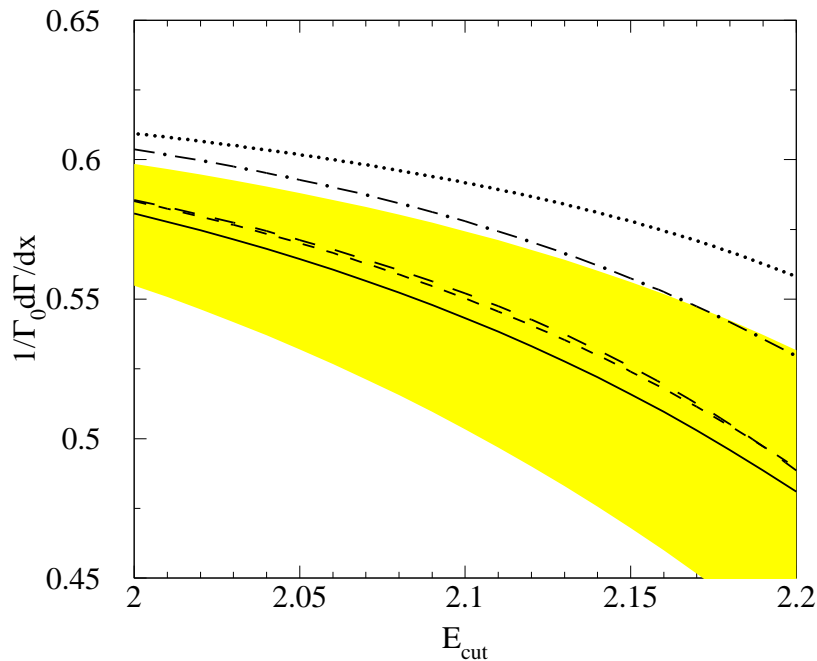} }
\tighten{
\caption[]{\it The rate integrated from $E_{cut}$.  The solid curve is
the resummed rate convoluted with the structure function in
Eq.~(\ref{SF1}), using the central values of $(\bar{\Lambda},\
\lambda_1)$.  The shaded region shows the uncertainty due to varying
the values of $(\bar{\Lambda},\ \lambda_1)$ by one sigma.  The dotted
curve is the one loop rate without the structure function.  The long
dashed curve is the resummed rate without the structure function.  The
dot-dashed curve is the one loop rate convoluted with the structure
function, using the central values of $(\bar{\Lambda},\ \lambda_1)$.
The short dashed curve is the one loop rate and the $O(\alpha_s^2)$
piece from the resummation, convoluted with the structure function.}}
\end{figure}
In Fig.~3 we show the rate as a function of the lower cut on the
photon energy, for the ansatz in Eq.~(\ref{SF1}) and with the central
values for $(\bar{\Lambda},\ \lambda_1)$.  In this plot we also show
the rate with and without perturbative resummation without the
inclusion of the structure function, and the one loop rate as well as
the one loop rate plus the $O(\alpha_s^2)$ piece from the expansion of
the resummed rate, convoluted with the structure function. As
expected, the effects of both perturbative and non-perturbative
resummation are enhanced as the photonic energy cut is increased.  We
see again that the dominant piece of the resummation is coming from
the $O(\alpha_s^2)$ term.  Furthermore, the inclusion of the structure
function does not significantly enhance the effects of the resummation
for these choices of structure functions. If they died off more slowly
for large negative $k_+$, we would expect the logs to be enhanced.
The shaded region is the resummed rate using the one sigma values for
$(\bar{\Lambda},\ \lambda_1)$, with again the structure function
defined in Eq.~(\ref{SF1}). We see that, as $\bar{\Lambda}$ is
increased, the effects of resummation become more important, as one
would expect since the width of the primordial distribution is
increasing. Varying the width $(\bar{\Lambda},\ \lambda_1)$, is one
way of determining the uncertainty due to our ignorance of the
structure function. Another method would be varying the functional
form of the structure function itself ({\it i.e.}~changing the higher
moments).

\begin{figure}[t]
\centerline{\epsfysize=11truecm  \epsfbox{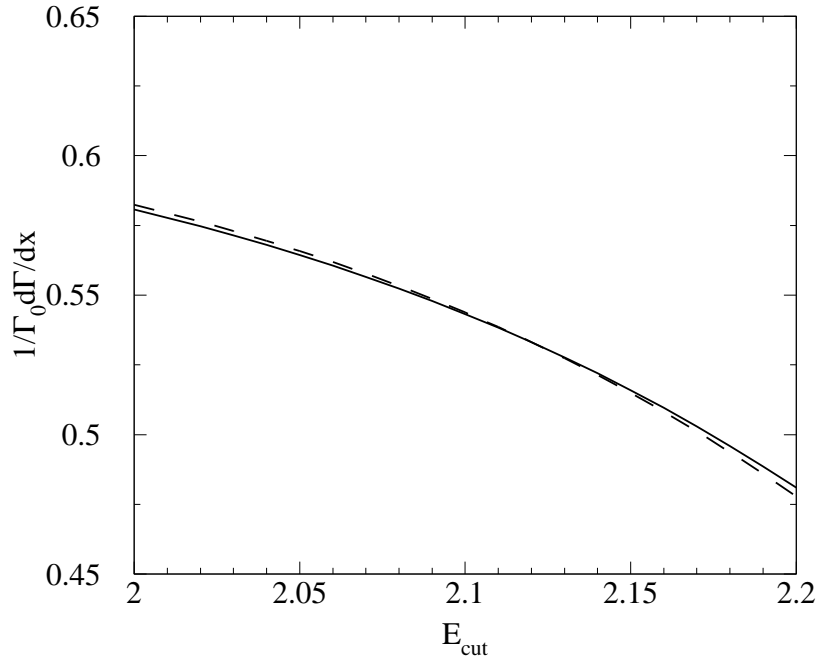} }
\tighten{
\caption[]{\it Rate convoluted with the different structure functions.
The solid curve is the rate convoluted with the Eq.~(\ref{SF2}), and the
dashed curve is the rate convoluted with Eq.~(\ref{SF2}).
}}
\end{figure}
In Fig.~4, we compare the rates for the two differing functional forms
given by Eq.~(\ref{SF1}) and Eq.~(\ref{SF2}).  We see that the rate is
not as sensitive to the higher moments as it is to the lower moments
$(\bar{\Lambda},\ \lambda_1)$. Note that for these particular
choices of the structure function the fourth and fifth moments differ by only
$O(50\%)$.  So perhaps a more thorough search of the possible form of
the structure functions should be explored.  Assuming that the
structure function is well behaved (read physically motivated), the
dominant uncertainty in the rate due to our ignorance of the structure
function can be removed once we have a better determination of
$(\bar{\Lambda},\ \lambda_1)$.

Finally, we may investigate the convergence of the expansion of $g_1$
in the exponent with the inclusion of the structure function.  In
Fig.~5 we show the rate for the series expansion of $g_1$ up to
$O(\alpha_s^3)$, we see that it is indeed well behaved and converges
rapidly. This again is signaling that the resummation is not
collecting a set of dominant terms.
\begin{figure}[t]
\centerline{\epsfysize=11truecm  \epsfbox{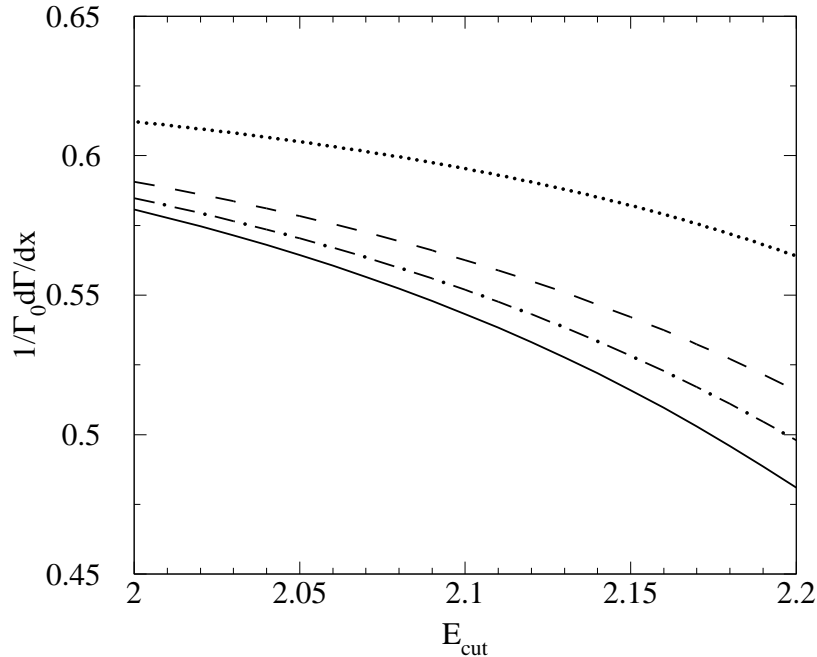} }
\tighten{
\caption[]{\it Rate convoluted with the structure function
Eq.(\ref{SF1}).  The dotted curve corresponds to expanding $g_1$ to
$O(\alpha_s)$, the dashed curve is for $g_1$ expanded to
$O(\alpha_s^2)$, and the dot-dashed curve is expanded to
$O(\alpha_s^3)$.  The solid curve is the rate with the full $g_1$.}}
\end{figure}

\section{Conclusions}

We conclude that resummation is not necessary, and does not increase
the accuracy of the prediction, when the photonic cut is near
$2.1{\rm\ GeV}$.  Therefore, for the purpose of calculating the total
rate, it is consistent to convolve the structure function with the
$O(\alpha_s$) partonic rate. The dominant errors will then come from
our ignorance of the structure function and higher order, in
$\alpha_s$, corrections to the rate, as well as uncertainties in
$\Gamma_0$ due to the dependence on the $b$ quark mass ({\it
i.e.}~$\bar\Lambda$).  We estimate that the uncertainties due to the
structure function are at the $20\%$ level, though we believe that a more
thorough search of the space of structure functions should be performed.
If we assume that the size of the $O(\alpha_s^2)$ corrections are
typically of the size of the pieces of the two loop result which we
pick off from our resummed rate, then the uncertainty coming from
these corrections is on the $10\%$ level.  On the other hand, if we
wish to extract the structure function from the measurement of the
spectrum for $b\rightarrow s\gamma$, in order to utilize it in the
extraction of $V_{ub}$, the resummation will most probably be
necessary.

\acknowledgments 
We wish to thank Tom Imbo, Ben Grinstein and Alex Kagan for useful
discussions.  We would like to thank Michelangelo Mangano, for
supplying us with a numerical check of our integrals, as well as
valuable discussions.  We also thank Adam Falk for his comments on this
manuscript.  This work was supported in part by the Department of
Energy under grant number DOE-ER-40682-143.

{\tighten

} 

\end{document}